\documentclass[11pt]{scrartcl}

\newcommand{\icpigheader}{28$^\text{th}$ ICPIG, July 15-20, 2007, Prague, Czech Republic}
\newcommand{\topicnumber}{B5}
\usepackage{cite}
\usepackage{amsmath,amssymb}
\usepackage{ulem}
\usepackage{graphicx,xcolor}
\usepackage[small]{caption2}
\usepackage[paper=a4paper,left=20mm,right=20mm,top=25mm,bottom=25mm]{geometry}
\newfont{\titelfont}{cmbx10 scaled 1400}
\newfont{\authorfont}{cmr10 scaled 1200}
\newfont{\institutefont}{cmsl10 scaled 1000}
\newfont{\abstractfont}{cmr10 scaled 1000}
\newfont{\captionfontICPIG}{cmr10 scaled 1000}

\newlength\abstractwidth
\newcommand{\presenter}[1]{\uline{#1}}

\makeatletter
	\renewcommand*\bib@heading{}
\makeatother

\renewenvironment{thebibliography}[1]
{
	\begin{oldthebibliography}{#1}
	\setlength{\parskip}{0ex}
	\setlength{\itemsep}{0ex}
}
{ 
	\end{oldthebibliography}
}
\usepackage{fancyhdr}
\graphicspath{{eps/}}
\begin{document}
\sloppy
\twocolumn[
\center{
{\titelfont Particle-based modeling of oxygen discharges}\\[0.5cm]
{\authorfont \presenter{F. X. Bronold}$^1$, K. Matyash$^2$, David Tskhakaya$^3$, 
             Ralf Schneider$^2$, and Holger Fehske$^1$}\\[0.5cm]
{\institutefont $^1$ Institut f\"ur Physik, Ernst-Moritz-Arndt-Universit\"at Greifswald, 
                D-17489 Greifswald}

{\institutefont $^2$ Max-Planck-Institut f\"ur Plasmaphysik, Teilinstitut Greifswald, 
                D-17491 Greifswald}

{\institutefont $^3$ Institut f\"ur Theoretische Physik, Universit\"at Innsbruck, 
                A-6020 Innsbruck, \"Osterreich}
}

\center{
\parbox[b]{\abstractwidth}{
\abstractfont
We present an one-dimensional particle-in-cell Monte-Carlo
model for capacitively coupled radio-frequency discharges in oxygen.
The model quantitatively describes the central part of the
discharge. For a given voltage and pressure, it self-consistently determines 
the electric potential and the distribution functions for electrons, negatively 
charged atomic oxygen, and positively charged molecular oxygen.
Previously used collision cross sections are critically 
assessed and in some cases modified. Provided associative detachment 
due to metastable oxygen molecules is included in the model, the 
electro-negativities in the center of the discharge are in excellent 
agreement with experiments. Due to lack of empirical data for the
cross section of this process, we propose a simple model and discuss its 
limitations. 
%At least 8, maximal 10 lines.
}
\vspace*{0.5cm}
}
]

{\noindent\textbf{1. Introduction}}\\
\indent
Discharges in reactive gases such as oxygen play an important role in 
plasma-assisted etching and thin-film deposition techniques. The requirements 
on the controllability and reliability of these discharges are so high, that 
further advancement of this technology critically depends on improved 
descriptions of the physical processes. In particular, the interplay between
the macroscopic electrodynamics, which is used to control the discharge, 
and the microscopic plasma-chemical processes, which sustain the discharge 
and give rise to the materials processing, has to be understood not only
qualitatively but quantitatively. 

Particle-based modeling is well suited for this task, 
because it directly simulates the Boltzmann-Poisson system describing 
the discharge without any assumptions concerning the species' distribution 
functions or the electric field. Provided the elementary plasma-chemical 
processes are well characterized in terms of cross sections, they can be 
easily incorporated in the collision integral of the Boltzmann equation. 
Through the source term of the Poisson equation, the plasma-chemistry is 
then linked to the electrodynamics of the discharge. 

In the following we give a brief account of our implementation of the
particle-in-cell Monte-Carlo collision (PIC-MCC) approach for the 
modeling of capacitively coupled radio-frequency (rf) discharges in
oxygen. Our main focus will be the critical assessment of cross section 
data for $({\rm O_2^+,O_2})$ charge exchange scattering, ion-ion neutralization, 
and detachment due to metastables. A full description of our approach, whose 
treatment of collisions is closely related to the direct simulation Monte-Carlo 
approach for rarefied gases, will be given elsewhere~\cite{BMT07}.\\ 

{\noindent\textbf{2. Model}}\\
\indent
The description of an oxygen discharge could be based on a brute force numerical 
solution of the Boltzmann-Poisson system which couples the distribution functions 
of the relevant species with the electric potential. In most cases, however, 
this approach is not practical.
More promising are methods which track the spatio-temporal evolution of a sample of 
pseudo-particles subject to elastic, inelastic, and reactive collisions. These approaches 
are based on the decoupling of collisions from the free flights in the self-consistent 
electric field. When the cross sections for the collisions are known, the durations 
of the free flights, as well as the probability for a collision of a particular type 
to occur, can be simply obtained from elementary kinetic considerations. 

\begin{figure}[t]
\center
\includegraphics[clip,width=0.70\linewidth]{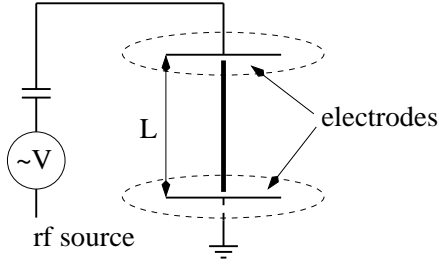}
\caption{Schematic geometry of the rf discharges used in Refs.~\cite{Dittmann07,KSQ00}.
We use an 1D model to simulate the central axial part of the
discharge (thick solid line).}
\label{geometry}
\end{figure}

Our simulations of capacitively coupled rf discharges in 
oxygen~\cite{Dittmann07,KSQ00} are restricted 
to the central axial part of the reactor (see Fig.~\ref{geometry}). 
Ignoring the (electric) asymmetry between the grounded and powered electrode, 
we use an one-dimensional (1D) model, which keeps only one spatial 
coordinate, $0\le x \le L$, where $L$ is the distance between the electrodes,
but retains all three velocity coordinates 
$v_x, v_y,$ and $v_z$. One of the electrodes is electrically driven by 
a time-dependent voltage, $U_{\rm rf}(t)=U\sin2\pi f$, while the
other is set to $U=0$. Both electrodes are totally absorbing, secondary 
electron emission is neglected, and the oxygen molecules are treated as 
an inexhaustible reservoir, the density of which is 
$n_{{\rm O_2}}=p/kT$, where $T=300~K$ and $p$ is the gas  pressure.

The complex plasma-chemistry of oxygen gives rise to a large variety of 
collisions. In Table~\ref{table_reac} we show the
ones with the largest cross sections. They are included in our model. We 
simulate only three species: electrons ($e$), negatively charged atomic 
oxygen (${\rm O}^-$), and positively charged molecular oxygen ($O_{2}^{+}$). 
Neutral particles 
%appearing either as educts or products 
are not treated kinetically. They are only 
accounted for in as far as their production leads to an energy loss for electrons 
(collisions~(10)--(15)) and in as far as they affect the balance of simulated
particles (reactions~(16)--(22)).

\begin{table}[t]%[H] add [H] placement to break table across pages
\caption{\label{reactions} Elastic, inelastic, and reactive collisions included 
in our model~\cite{BMT07}. 
%A detailed discussion of it as well as a critical assessment of 
%the cross sections used to characterize the collisions will be given elsewhere~\cite{BMT07}.
}
\begin{tabular}{ll}
\hline\hline
elastic scattering                                      & \\
(1) ${e + e \rightarrow e + e}$                         & \\
(2) ${\rm O^- + O^- \rightarrow O^- + O^-}$             & \\
(3) ${\rm O_2^+ + O_2^+ \rightarrow O_2^+ + O_2^+}$     & \\
(4) $e + {\rm O_2^+} \rightarrow e + {\rm O_2^+}$       & \\
(5) $e + {\rm O^-}   \rightarrow e + {\rm O^-}$         & \\
(6) ${\rm O^- + O_2^+ \rightarrow O^- + O_2^+}$         & \\
(7) ${\rm e + O_2 \rightarrow e + O_2}$                 & \\
(8) ${\rm O^- + O_2 \rightarrow O^- + O_2}$             & \\
(9) ${\rm O_2^+ + O_2 \rightarrow O_2 + O_2^+}$         & \\
electron energy loss scattering                         & \\
(10) ${\rm e + O_2 \rightarrow e + O_2(\nu=1,...,4)}$   & \\
(11) ${\rm e + O_2 \rightarrow e + O_2(Ryd)}$           & \\
(12) ${\rm e + O_2 \rightarrow e + O(3P) + O(3P)}$      & (6.4~eV)\\
(13) ${\rm e + O_2 \rightarrow e + O(3P) + O(1D)}$      & (8.6~eV)\\
(14) ${\rm e + O_2 \rightarrow e + O_2(a^1\Delta_g)}$   & \\
(15) ${\rm e + O_2 \rightarrow e + O_2(b^1\Sigma_g)}$   & \\
electron $\&$ ion production $\&$ loss                  & \\
(16) ${\rm e + O_2^+ \rightarrow O + O}$                & \\
(17) ${\rm O^- + O_2^+ \rightarrow O + O_2}$            & \\
(18) ${\rm e + O_2 \rightarrow O + O^-}$                & \\
(19) ${\rm O^- + O_2 \rightarrow O + O_2 + e}$          & \\
(20) ${\rm O^- + O_2(a^1\Delta_g) \rightarrow O_3 + e}$ & \\
(21) ${\rm e + O_2 \rightarrow 2e + O_2^+}$             & \\
(22) ${\rm e + O^- \rightarrow O + 2e}$                 & \\
\hline\hline
\end{tabular}
\label{table_reac}
\end{table}

\begin{figure}[t]
\center
\includegraphics[clip,width=0.9\linewidth]{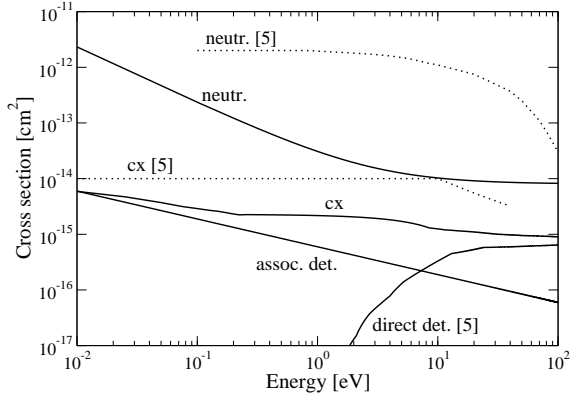}
\caption{Cross sections for $({\rm O_2^+, O_2})$ charge exchange scattering (9),
ion-ion neutralization (17), direct detachment (19), and associative (20) 
detachment. The dotted lines indicate the respective cross sections used in 
Ref.~\cite{VS95}.} 
\label{Xsection}
\end{figure}

Our collection of cross sections is semi-empirical, combining measured data
with models for the low-energy asymptotic. A complete discussion 
of the molecular physics entering the simulation will be given elsewhere~\cite{BMT07}. 
Here it suffices to mention that the cross sections for $({\rm O_2,O_2^+})$ charge 
exchange scattering (9), ion-ion neutralization (17), and detachment (19,20) significantly 
deviate from the ones used previously~\cite{VS95} (see Fig.~\ref{Xsection}). Our simulations 
indicate that the modifications are essential for obtaining results in 
accordance with experiments~\cite{Dittmann07,KSQ00}. 

Using $\sigma_{cx}(E)=\sigma_m(E)/2$,
where $\sigma_m$ and $\sigma_{cx}$ denote, respectively, the momentum and
charge exchange cross section, and $E$ is the relative kinetic energy, we based 
our cross section for $({\rm O_2,O_2^+})$ charge
exchange scattering below $0.251~eV$ and above $8.5~eV$ on empirical data for momentum 
scattering~\cite{STS63,GR72}. For energies in between, we employed a
linear interpolation. With this cross section, we obtained ${\rm O}_2^+$ velocity 
distribution functions in close agreement with experiments~\cite{Dittmann07}. 
With the charge exchange cross section given in Ref.~\cite{VS95}, on the other 
hand, we could not reproduce the experimental findings -- neither with our 
PIC-MCC code nor with the BIT1 code~\cite{TK02}, which we used to cross-check our results.

For ion-ion neutralization, we employed the cross section of a two-channel
Landau-Zener model~\cite{Olson72}, $\sigma_{n}(E) = 4\pi R_x^2\big(1+\frac{1}{R_x E}\big)$,
where $R_x$ is a free parameter which we adjusted to obtain the 
experimentally measured cross section at high energies~\cite{PP98}. This 
cross section deviates dramatically from the one used in Ref.~\cite{VS95}, but it 
is based on a clear physical picture for the neutralization process as well as empirical 
data at high energies. 

Detachment of ${\rm O}^-$ on neutrals occurs through direct detachment (19) 
and associative detachment (20). The latter is rather surprising because 
there is no evidence for it in beam experiments~\cite{CS74a}. Yet, investigations 
of ${\rm O}_2$ discharges strongly suggest that associative detachment is 
possible because of the presence of metastable ${\rm O}_2(a^1\Delta_g)$ and 
may be even the main loss process for ${\rm O}^-$ in some 
pressure range~\cite{KSQ00}. Due to lack of empirical data for this process, 
we employed a simple model, which describes detachment as the ``inverse'' of a 
Langevin-type electron capture into an attractive auto-detaching state of 
${\rm O}_3^-$. The cross section of which reads 
\begin{eqnarray}
\sigma^\Delta_{ad}(E) =
5.96 \cdot \frac{10^{-16}\cdot cm^2}{\sqrt{E[eV]}}~,
\label{Xsection_ad}
\end{eqnarray}
where we assumed the polarizability of ${\rm O}_2(a^1\Delta_g)$ to be the same
as for ${\rm O_2}$. Note, in contrast to the cross section for direct detachment,
the cross section for associative detachment has no threshold. To determine 
the probability for this process, 
we also need the density of ${\rm O_2(a^1\Delta_g)}$. Within the three species
model this density is unknown. It should be however of the order of the 
${\rm O}_2$ density. As a first step, we write therefore $n_\Delta = C\cdot n_{{\rm O}_2}$,
with $C<1$ an adjustable fit parameter.\\

{\noindent\textbf{3. Results}}\\
\indent
Similar to other electro-negative gas discharges, the presence of
negative ions in an oxygen discharge leads to abrupt changes in the
ion density which, in most cases, forces the discharge to stratify into a
quasi-neutral ion-ion and a peripheral electro-positive edge plasma. The
details of the stratification depend on the interplay between
plasma-chemistry and electrodynamics. 

In oxygen discharges two plasma-chemical processes are of particular 
importance: Ion-ion neutralization and associative detachment due 
${\rm O_2(a^1\Delta_g)}$. Although the three species plasma model 
cannot fully describe associative detachment, because it assumes an
homogeneous background of ${\rm O_2(a^1\Delta_g)}$ molecules,
it nevertheless gives clear evidence that the latter process 
is indispensable for a correct description of experiments. 

\begin{figure}[t]
\center
\includegraphics[clip,width=0.9\linewidth]{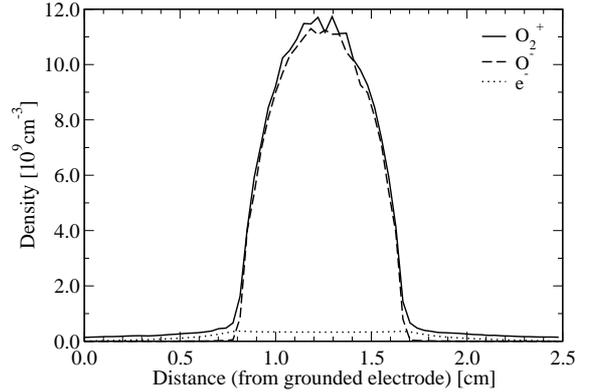}
\caption{Cycle-averaged electron and ion density profiles for a rf discharge
in ${\rm O}_2$ without associative detachment (20) taken into account. The
parameters of the discharge are $L=2.5~cm$, $f=13.6~MHz$, $p=13.8~Pa$
and $U=250~V$.}
\label{without}
\end{figure}

\begin{figure}[t]
\center
\includegraphics[clip,width=0.9\linewidth]{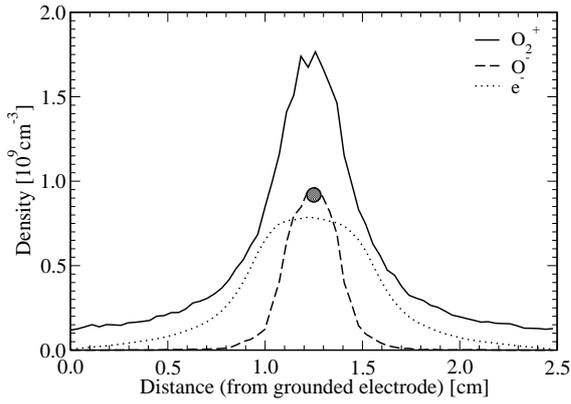}
\caption{Same plot as Fig.~\ref{without} but now with associative detachment~(20)
taken into account. 
%The parameters of the discharge are unchanged. 
Experimentally, $n_e\approx n_{{\rm O^-}}\approx 0.9\cdot 10^9~cm^{-3}$
(grey bullet)~\cite{KSQ00}.} 
\label{with}
\end{figure}

To demonstrate this, we simulated the discharge of Katsch and coworkers~\cite{KSQ00}. 
In Figs.~\ref{without} and \ref{with} we show, respectively, the quasi-stationary, 
cycle-averaged density profiles of the simulated 
charged particles for $p=13.8~Pa$, $U=250~V$, $L=2.5~cm$, and 
$f=13.6~MHz$ without and with associative detachment taken 
into account. The parameter $C \approx 1/6$, implying that one-sixth
of the ${\rm O}_2$ molecules are in the metastable state.
% is here tuned to approximately 
%reproduce the experimental densities in the center of the discharge. 
The precise value of $C$ should not be taken too serious because it is
based on a rather crude model for associative detachment. 
More important is that without this process ($C=0$), the simulation
could not reproduce the measured densities. 

Whereas the simulation with associative detachment reproduces reasonably 
well the densities of charged particles in the center of
the discharge, the shapes of the (axial) density profiles deviate from 
the measured ones. Compared to experiment, the central plasma is 
too narrow, most notably, for lower voltages (not shown here)~\cite{BMT07}. 
This is a shortcoming of the three species plasma model which ignores the 
spatial dependence of the ${\rm O_2(a^1\Delta_g)}$ density which, in 
reality, results from the interplay of volume and surface loss and generation 
processes. Because the probability of associative detachment is
proportional to $n_{\Delta}$, the ${\rm O_2(a^1\Delta_g)}$ density profile
should strongly affect the density profiles of charged particles.

\begin{figure}[t]
\center
\includegraphics[clip,width=0.9\linewidth]{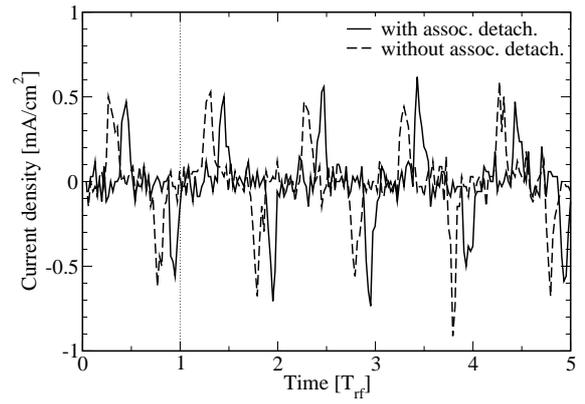}
\caption{Time dependence (for five rf cycles with duration $T_{\rm rf}$) of the current 
densities corresponding 
to the cycle-averaged, quasi-stationary density profiles shown in Figs.~\ref{without} 
and~\ref{with}, respectively.}
\label{currents}
\end{figure}

In Fig.~\ref{currents} we finally plot the time-resolved current densities through 
the discharges shown in Figs.~\ref{without} and \ref{with}. The current densities
are almost the same, irrespective of whether associative detachment is 
taken into account or not. This is a consequence of the fact that in both cases 
the main part of the current is carried by the electrons whose densities, in turn, 
are basically identical. Further studies are needed to reveal if this is 
an artifact of our model, or if 
%-- for an oxygen discharge -- 
indeed two rather 
different density profiles are consistent with a given external power supply,
$\langle U_{\rm rf}\cdot j_{\rm rf}\rangle_{\rm rf} \cdot A$, where $A$ is the 
area of the electrodes and $\langle...\rangle_{\rm rf}$ denotes the cycle 
average. The configuration realized in the discharge depends then on the 
plasma-chemistry, in particular, on the outcome of the competition between 
ion-ion neutralization and associative detachment.\\ 

%\begin{figure}[ht]
%\centering
%\includegraphics[clip,width=7cm]{PhaseSensitive}
%\caption{Representative phase-sensitive measurements and simulations.}
%\label{fig:phase}
%\end{figure}

{\noindent\textbf{4. Conclusions}}\\
\indent
We constructed an 1D three-species PIC-MCC model for capacitively coupled 
rf discharges in oxygen. The model simulates three kinds of charged 
particles ($e$, ${\rm O^-}$, and ${\rm O_2^+}$) and retains neutral 
particles indirectly via collisions with the simulated charged 
particles. This is sufficient to reproduce measured central electron and 
ion densities. However, the (axial) ion density profiles of the simulations
are too narrow compared to the experimental ones. We expect better 
agreement, when the modeling also allows for the possibility of 
${\rm O}_2(a^1\Delta_g)$ density profiles.
For that purpose, the ${\rm O}_2(a^1\Delta_g)$ molecules,
together with their main loss and generation processes, 
have to be explicitly included in the Monte-Carlo collision approach.\\

Support from the SFB-TR 24 ``Complex Plasmas'' is greatly acknowledged.
We thank B. Bruhn, H. Deutsch, K. Dittmann, and J. Meichsner for 
valuable discussions.
K. M. and R. S. acknowledge funding of the work by the Initiative and 
Networking Fund of the Helmholtz Association.\\

{\noindent\textbf{References}}
\begin{flushleft}

\end{flushleft}


\begin{thebibliography}{10}
\providecommand{\url}[1]{\texttt{#1}}
\providecommand{\urlprefix}{URL }

\bibitem{BMT07}
F.~X. Bronold {\it et al.}, unpublished.

\bibitem{Dittmann07}
K.~Dittmann {\it et al.}, unpublished.

\bibitem{KSQ00}
H.~M. Katsch {\it et al.}, Plasma Sources Sci. Technol.
  \textbf{9} (2000) 323.

\bibitem{VS95}
V.~Vahedi and M.~Surendra, Comput. Phys. Commun. \textbf{87} (1995) 179.

\bibitem{STS63}
R.~F. Stebbings {\it et al.}, J. Chem. Phys. \textbf{38}
  (1963) 2277.

\bibitem{GR72}
D.~R. Gray and J.~A. Rees, J. Phys. B \textbf{5} (1972) 1048.

\bibitem{TK02}
D.~Tskhakaya and S.~Kuhn, Contrib. Plasma Phys. \textbf{42} (2002) 302.

\bibitem{Olson72}
R.~E. Olson, J. Chem. Phys. \textbf{56} (1972) 2979.

\bibitem{PP98}
R.~Padgett and B.~Peart, J. Phys. B \textbf{31} (1998) L995.

\bibitem{CS74a}
J.~Comer and G.~J. Schulz, J. Phys. B \textbf{7} (1974) L249.

\end{thebibliography}
\end{document}